\documentclass[]{aa}
\usepackage{graphicx}
\begin{document}
%
\title{Evidence for environment-dependent galaxy Luminosity Function
up to $z=1.5$ in the VIMOS-VLT Deep Survey\thanks{based on data obtained 
with the European Southern Observatory Very Large
    Telescope, Paranal, Chile, program 070.A-9007(A), and on data
    obtained at the Canada-France-Hawaii Telescope, operated by the
    CNRS of France, CNRC in Canada and the University of Hawaii.}\\
}

\author{
O. Ilbert \inst{1}
\and O. Cucciati \inst{2}
\and C. Marinoni \inst{3,4}
\and L. Tresse \inst{4}
\and O. Le F\`evre \inst{4}
\and G. Zamorani \inst{5}
\and S. Bardelli  \inst{5}
\and A. Iovino \inst{2}
\and E. Zucca    \inst{5}
\and S. Arnouts \inst{4}
\and D. Bottini \inst{6}
\and B. Garilli \inst{6}
\and V. Le Brun \inst{4}
\and D. Maccagni \inst{6}
\and J.P. Picat \inst{8}
\and R. Scaramella \inst{12}
\and M. Scodeggio \inst{6}
\and G. Vettolani \inst{7}
\and A. Zanichelli \inst{7}
\and C. Adami \inst{4}
\and M. Bolzonella  \inst{1} 
\and A. Cappi    \inst{5}
\and S. Charlot \inst{11}
\and P. Ciliegi    \inst{5}  
\and T. Contini \inst{8}
\and S. Foucaud \inst{6}
\and P. Franzetti \inst{6}
\and I. Gavignaud \inst{8,13}
\and L. Guzzo \inst{2}
\and B. Marano     \inst{1}  
\and A. Mazure \inst{4}
\and H.J. McCracken \inst{11}
\and B. Meneux \inst{6}
\and R. Merighi   \inst{5} 
\and S. Paltani \inst{14,15}
\and R. Pell\`o \inst{8}
\and A. Pollo \inst{4}
\and L. Pozzetti    \inst{5} 
\and M. Radovich \inst{9}
\and M. Bondi \inst{7}
\and A. Bongiorno \inst{1}
\and G. Busarello \inst{9}
\and S. De La Torre \inst{4}
\and L. Gregorini \inst{10}
\and F. Lamareille \inst{8}
\and G. Mathez \inst{8}
\and Y. Mellier \inst{11,16}
\and P. Merluzzi \inst{9}
\and V. Ripepi \inst{9}
\and D. Rizzo \inst{8}
\and D. Vergani \inst{6}
}

\offprints{O.~Ilbert, e-mail: olivier.ilbert1@bo.astro.it}
   
\institute{
Universit\`a di Bologna, Dipartimento di Astronomia - Via Ranzani 1, 40127, Bologna, Italy
\and
INAF-Osservatorio Astronomico di Brera - Via Brera 28, Milan, Italy
\and
Centre de Physique Th\'eorique, Marseille, France
\and
Laboratoire d'Astrophysique de Marseille, UMR 6110 CNRS-Universit\'e de Provence, BP8, 13376 Marseille Cedex 12, France
\and
INAF-Osservatorio Astronomico di Bologna - Via Ranzani 1, 40127, Bologna, Italy
\and
IASF-INAF - via Bassini 15, 20133, Milano, Italy
\and
IRA-INAF - Via Gobetti,101, 40129, Bologna, Italy
\and
Laboratoire d'Astrophysique de l'Observatoire Midi-Pyr\'en\'ees, UMR 5572, 14 avenue E. Belin, 31400 Toulouse, France
\and
INAF-Osservatorio Astronomico di Capodimonte - Via Moiariello 16, 80131, Napoli, Italy
\and
INAF-Osservatorio Astronomico di Brera - Via Brera 28, Milan, Italy
\and
Institut d'Astrophysique de Paris, UMR 7095, 98 bis Bvd Arago, 75014 Paris, France
\and
INAF-Osservatorio Astronomico di Roma - Via di Frascati 33, 00040, Monte Porzio Catone, Italy
\and
European Southern Observatory, Garching, Germany
\and
Integral Science Data Centre, ch. d'\'Ecogia 16, CH-1290 Versoix
\and
Geneva Observatory, ch. des Maillettes 51, CH-1290 Sauverny
\and
Observatoire de Paris, LERMA, 61 Avenue de l'Observatoire, 75014 Paris, 
France}

\date{Received ... / Accepted ... }
\titlerunning{Environment-dependent galaxy Luminosity Function
up to $z=1.5$}
\authorrunning{Ilbert et al.}

\abstract{ We measure the evolution of the galaxy Luminosity Function
  as a function of large-scale environment up to $z=1.5$ from the
  VIMOS-VLT Deep Survey (VVDS) first epoch data. The 3D galaxy density
  field is reconstructed using a gaussian filter of smoothing length
  $5 h^{-1}$ Mpc and a sample of 6582 galaxies with $17.5 \le I_{AB}
  \le 24$ and measured spectroscopic redshifts. We split the sample in
  four redshift bins up to $z=1.5$ and in under-dense and over-dense
  environments according to the average density contrast $\delta=0$.
  There is a strong dependence of the Luminosity Function (LF) with
  large-scale environment up to $z=1.2$: the LF shape is observed to
  have a steeper slope in under-dense environments. We find
  $\alpha=-1.32\pm 0.07, -1.35\pm 0.10, -1.42\pm 0.18$ in under-dense
  environments and $\alpha=-1.08\pm 0.05, -1.06\pm 0.06$, $-1.22\pm
  0.12$ in over-dense environments in the redshift bins
  $z=$[$0.25-0.6$], [$0.6-0.9$], [$0.9-1.2$], respectively using a
  best-fit Schechter luminosity function.  We find a continuous
  brightening of $\Delta M^* \sim 0.6$ mag from $z=0.25$ to $z=1.5$
  both in under-dense and over-dense environments.  The rest-frame
  $B$-band luminosity density continuously increases in under-dense
  environments from $z=0.25$ to $z=1.5$ whereas its evolution in
  over-dense environments presents a peak at $z\sim 0.9$.  We interpret
  the peak by a complex interplay between the decrease of the star
  formation rate and the increasing fraction of galaxies at $\delta>0$
  due to hierarchical growth of structures.
  As the environmental dependency of the LF shape is
  already present at least up to $z=1.2$, we therefore conclude that
  either the shape of the LF is imprinted very early on in the life of
  the Universe, a `nature' process, or that `nurture' physical
  processes shaping up environment relation have already been
  efficient earlier than a look-back time corresponding to 30\% of the
  current age of the Universe.

   }

   \maketitle
%

\section{Introduction}

Environmental effects are expected to play a key role in galaxy
formation and evolution. A tight morphology-density relationship
between galaxy morphology and local environment has been firmly
established in clusters of galaxies, showing a strong decrease of the
spiral population in the densest environments to the benefit of
early-type galaxies (e.g. \cite{Dressler80}1980). In the last few
years, environmental effects have been systematically investigated in
less extreme density regimes. In particular the Sloan Digital Sky
Survey (SDSS; \cite{York00}2000) and the 2dF Galaxy Redshift Survey
(2dFGRS; \cite{Colless01}2001) have allowed a complete census of the
large-scale galactic environment in the local universe. At $z\sim 0.1$
and over a wide range of local galaxy densities, a strong dependency
of galaxy properties with environment is observed; it includes a
systematic decrease of the star formation rate towards over-dense
regions (e.g.  \cite{Kauffmann04} 2004, \cite{Gomez03} 2003) and a
larger population of red galaxies in over-dense regions
(\cite{Balogh04} 2004).

Particular attention has also been devoted to investigate eventual
environmental imprints on the galaxy Luminosity Function (LF).
Environmental studies using small local redshift samples have provided
evidence that the LF in dense environments shows a brighter $M^*$ for
groups of galaxies with increasing richness, whereas the slope does
not change significantly from the field to groups (\cite{Marinoni99}
1999, \cite{Ramella99} 1999).  The analysis based on the large SDSS
and 2dFGRS surveys has confirmed this picture. For example, at
$0.05<z<0.13$, \cite{Croton05}(2005) find that the LF is sensitive to
the large-scale environment with a top-hat smoothing length at $8
h^{-1}$Mpc. They show that the galaxy LF brightens continuously from
voids to clusters with no significant variation of the LF
slope. Studies dedicated to specific environments go in the same
direction: \cite{Hoyle05}(2005) find a fading of the LF in voids
compared to field galaxies in the SDSS; \cite{DePropris02}(2002) find
a brightening of the LF with a sample of 60 clusters in the
2dFGRS. These measurements provide a highly constrained local
benchmark against which to compare environmental studies at higher
redshift.

The various physical and astrophysical mechanisms which are likely to
imprint an environmental dependency on galaxy properties may be
classified as `nurture' or `nature' processes. The first are expected
to be active over most of the life of a galaxy, whereas `nature' processes
would imprint an environmental dependency in some galaxy properties
early on in the galaxy evolution history (see
e.g. \cite{Kauffmann04} 2004).
We may discriminate inner galaxy properties from those which might be
sensitive to external environments, and progress toward a better
understanding of whether the dependence on environment was established
early on or developed gradually as a function of time, by directly
tracking the evolution of the relationship between galaxies and their
surrounding environment across different cosmic epochs.

In this paper we are analysing the dependency of the galaxy luminosity
function on environment up to a redshift z=1.5 from 6582 galaxies with
spectroscopic redshifts in the VIMOS-VLT Deep Survey (VVDS;
\cite{LeFevre05a}2005). We have reconstructed the 3D density field up
to $z=1.5$ using a gaussian filter with a smoothing scale of $5
h^{-1}$ Mpc in \cite{Cucciati06}(2006). Here we investigate the
dependency of the luminosity function on the large-scale environment
from $z=0.25$ to $z=1.5$. This paper is organized as follows: we
describe the data in \S 2, the environmental estimator in \S 3 and the
method to compute the LF in \S 4. In \S 5 we present our results on
the dependency of the LF and luminosity density on environment through
cosmic time.  Results are discussed in \S 6 and a summary is presented
in \S 7. We adopt a flat, vacuum dominated, cosmology
($\Omega_m~=~0.3$, $\Omega_\Lambda~=~0.7$) and we define
$h~=~\rm{H}_{\rm0}/100$~km~s$^{-1}$~Mpc$^{-1}$. Magnitudes are given
in the AB system.


\section{Data description}

We use the first epoch VVDS deep sample covering $1750$~arcmin$^2$
over the VVDS-0226-04 field (\cite{LeFevre05a}2005). The spectroscopic
targets are purely magnitude selected with $17.5 \le I_{AB} \le
24.0$. The sample consists of 6582 galaxies, 623 stars and 62 QSOs
with reliable spectroscopic measurements (confidence level from 80 to
100\%), 1439 objects with an uncertain redshift measurement
(confidence level of $\sim$55\%) and 690 objects with no spectroscopic
identification. In this paper we only consider galaxies with a
redshift measurement confidence level higher than 80\%. The median
redshift of the sample is 0.76 about with a 1$\sigma$ accuracy of the
redshift measurements estimated to be $dz \sim 0.0009$.

Deep multi-color images cover the entire area targeted in
spectroscopy.  $B$, $V$, $R$, $I$ photometry was acquired with the
wide-field 12K mosaic camera at the CFHT (McCracken et al. 2003;
\cite{LeFevre04}2004), $u^*$, $g'$, $r'$, $i'$, $z'$ photometry from
the Canada-France-Hawaii Legacy Survey ({\it
www.cfht.hawaii.edu/Science/CFHLS}).
The apparent magnitudes are Kron-like elliptical aperture magnitudes
(\cite{Kron80}1980) which have been corrected for galactic extinction
from dust map images (\cite{Schlegel98}1998).

\section{Large-scale environment: local galaxy density}

We briefly summarize the method detailed in \cite{Cucciati06}(2006) to
provide a quantitative measurement of the large-scale environment. 

The average density $<\rho>$ is estimated in a redshift slice with a
depth of $800 h^{-1}$Mpc centered on the galaxy $i$. We use a gaussian
smoothing method to dilute the observed point-like spatial
distribution with $R_S$ being the typical smoothing length.  A local
density contrast $\delta_i$ with respect to the average density
$<\rho>$ is associated to each galaxy $i$:
\begin{equation}
\delta_i(R_S)=\frac{\rho_i(R_S)~-~<\rho>}{<\rho> }.
\end{equation}
Each galaxy $j$ contributes to the local density $\rho_i$
according to:
\begin{equation}
\rho(r_i,~R_S)=\sum_{j \ne i} \frac{w_j}{S_j}\frac{1}{(2\pi
R_S^2)^{3/2}}e^{-\frac{(r_i-r_j)^2}{2R_S^2}}, 
\end{equation}
where $r$ is the distance between the galaxies $i$ and $j$, $w_j$ is
the weight associated to the galaxy $j$ to correct for the
spectroscopic incompleteness and $S_j$ is the selection function of a
flux-limited sample. Based on numerical simulations which mimic the
VVDS survey, \cite{Cucciati06}(2006) show that $R_S=5 h^{-1}$Mpc is
the smallest smoothing length for which the density contrast is well
recovered given the VVDS survey strategy and redshift sampling rate.
In the following we use $R_S=5 h^{-1}$Mpc and the threshold $\delta=0$
to separate under- and over-dense regions.

\section{Procedure to compute the Luminosity Function}

We derive the LF using the Algorithm for Luminosity Function (ALF)
described in Appendix A of \cite{Ilbert05}(2005). We measure
k-corrections from a procedure of template fitting on the multi-color
data. We derive the rest-frame luminosity at $\lambda$ from the
apparent magnitude observed at $\lambda \times (1+z)$ to minimize
k-correction uncertainties. Galaxies with different spectral type are
not visible up to the same faint absolute magnitude limit due to the
type dependency of the k-correction. This may strongly bias the LF
estimate (see \cite{Ilbert04}2004). We therefore restrict our estimate
to the absolute magnitude range complete in term of spectral type. We
present results obtained on the basis of the STY
(\cite{Sandage79}1979) and 1/V$_{\rm max}$ (\cite{Schmidt68}1968) LF
estimators. For each estimator we introduce a statistical weight
function of apparent magnitude and redshift that corrects for sources
not observed ({\it Target Sampling Rate}) and for which the
spectroscopic measurement failed ({\it Spectroscopic Success Rate})
(see \cite{Ilbert05}2005). The principle of the STY is to determine
the Schechter parameters which maximize the likelihood to observe a
given galaxy sample. This estimator allows us to describe the LF using
the three parameters $\alpha$, $M^*$ and $\phi^*$, and quantitatively
describe the LF evolution. The 1/V$_{\rm max}$ is complementary since
this non-parametric method does not presuppose any functional form for
the LF. Note that for both estimators, we are not considering the
volume associated to a specific environment as done by
\cite{Croton05}(2005) but the whole volume surveyed in the given
redshift bin.

\begin{figure*}
\centering \includegraphics[width=16cm]{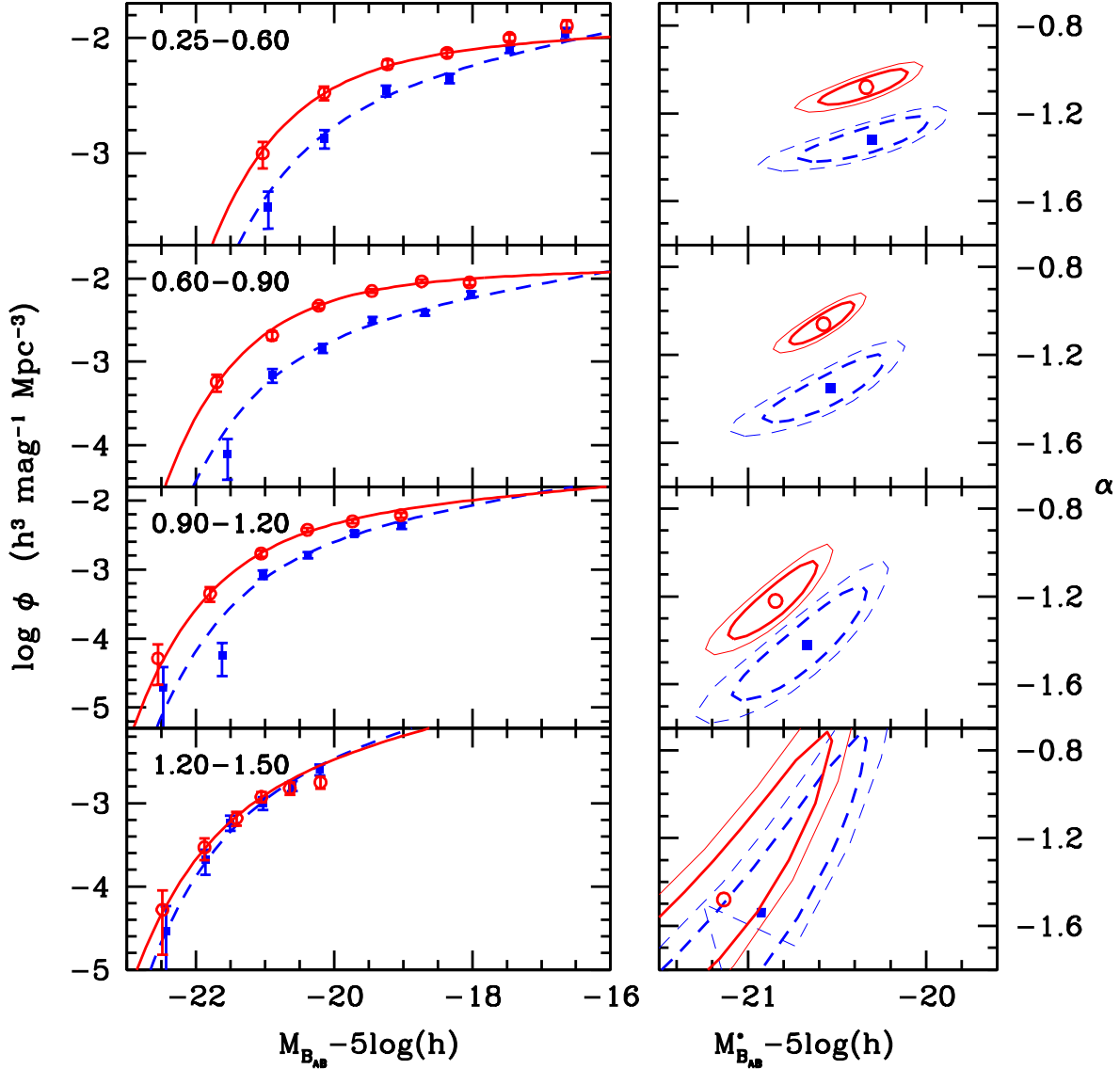}
\caption{Left panels: B-band LF for under-dense (blue solid squares
  and dashed lines) and over-dense environment (red open circles and
  solid lines) from $z=0.25-0.6$ (top panel) up to $z=1.2-1.5$ (bottom
  panel). The lines correspond to the best-fitting Schechter functions
  obtained with the STY estimator. The points correspond to the
  non-parametric estimate obtained with the 1/V$_{\rm max}$
  estimator. Right panels: the corresponding 68\% (thick lines) and
  90\% (thin lines) error contours obtained with the STY estimator.}
\label{LFglob}
\end{figure*}

\begin{table*}
\begin{center}
\begin{tabular}{ c c c c c c c } \hline
  &  & & & &  \multicolumn{1}{c}{$\Phi^*$}   & log $\rho_L(B)$   \\
 Environment &      z-bin &        N     &   $\alpha$ &  $M^*_{B_{AB}}-5log(h)$                    & ($10^{-3} h^3 Mpc^{-3}$)  & ($W Hz^{-1} Mpc^{-3}$) \vspace{0.2cm} \\ \hline
\hline
under-dense & 0.25-0.60 &   643 &  -1.32$^{{\rm + 0.07}}_{{\rm - 0.07}}$ & -20.31$^{{\rm + 0.22}}_{{\rm - 0.24}}$ &   3.59$^{{\rm + 0.92}}_{{\rm - 0.82}}$ & 19.302$^{\rm +  0.073}_{{\rm - 0.051}}$ \\
            & 0.60-0.90 &   641 &  -1.35$^{{\rm + 0.10}}_{{\rm - 0.10}}$ & -20.54$^{{\rm + 0.21}}_{{\rm - 0.22}}$ &   3.07$^{{\rm + 0.87}}_{{\rm - 0.78}}$ & 19.347$^{\rm +  0.069}_{{\rm - 0.045}}$ \\
            & 0.90-1.20 &   471 &  -1.42$^{{\rm + 0.18}}_{{\rm - 0.18}}$ & -20.67$^{{\rm + 0.23}}_{{\rm - 0.25}}$ &   3.60$^{{\rm + 1.31}}_{{\rm - 1.14}}$ & 19.511$^{\rm +  0.256}_{{\rm - 0.094}}$ \\
            & 1.20-1.50 &   193 &  -1.54$^{{\rm + 0.54}}_{{\rm - 0.55}}$ & -20.93$^{{\rm + 0.42}}_{{\rm - 0.52}}$ &   3.58$^{{\rm + 2.59}}_{{\rm - 2.38}}$ & 19.716$^{\rm +  23.833}_{{\rm - 0.338}}$ \\
over-dense & 0.25-0.60 &   924 &  -1.08$^{{\rm + 0.05}}_{{\rm - 0.05}}$ & -20.33$^{{\rm + 0.15}}_{{\rm - 0.16}}$ &   8.20$^{{\rm + 1.24}}_{{\rm - 1.17}}$ & 19.571$^{\rm +  0.059}_{{\rm - 0.046}}$  \\
           & 0.60-0.90 &  1440 &  -1.06$^{{\rm + 0.06}}_{{\rm - 0.06}}$ & -20.57$^{{\rm + 0.12}}_{{\rm - 0.12}}$ &  10.40$^{{\rm + 1.32}}_{{\rm - 1.28}}$ & 19.764$^{\rm +  0.030}_{{\rm - 0.026}}$  \\
           & 0.90-1.20 &   820 &  -1.22$^{{\rm + 0.12}}_{{\rm - 0.12}}$ & -20.84$^{{\rm + 0.16}}_{{\rm - 0.16}}$ &   6.63$^{{\rm + 1.38}}_{{\rm - 1.30}}$ & 19.735$^{\rm +  0.067}_{{\rm - 0.041}}$  \\
           & 1.20-1.50 &   194 &  -1.48$^{{\rm + 0.50}}_{{\rm - 0.51}}$ & -21.13$^{{\rm + 0.43}}_{{\rm - 0.53}}$ &   3.12$^{{\rm + 2.27}}_{{\rm - 2.06}}$ & 19.683$^{\rm +  23.798}_{{\rm - 0.271}}$  \\
\hline
\end{tabular}
\caption{Schechter parameters and associated one sigma errors
  ($2\Delta ln\mathcal L=1$) for the rest-frame $B$-band LF estimated
  with the STY. Parameters are given for under-dense and over-dense
  environments.}
\label{tab1}
\end{center}
\end{table*}

\section{The environment dependent Luminosity Function and Luminosity Density}

We split the sample in four redshift slices from $z=0.25$ to $1.5$ and
in two large-scale environment classes according to the density
contrast threshold $\delta=0$. The left panels of Fig.\,\ref{LFglob}
show the resulting LFs for under- and over-dense environments. We find
a clear dependence of the LF on galactic environment up to $z=1.2$.
From the non-parametric 1/V$_{\rm max}$ estimator, we observe an
excess of bright galaxies in over-dense environments compared to the
luminosity distribution of galaxies in low-density regions which
exhibits a larger number of faint galaxies.

We investigate this environmental dependency more quantitatively using
the STY best-fitting parameters. The right panels of
Fig.\,\ref{LFglob} show the STY best-fitting parameters with their
68\% and 90\% error contours. The best-fitting parameters are listed
in Table \ref{tab1}. We find that the LF shape depends strongly on the
large-scale environment: the differences between the $\alpha-M^*$
best-fitting values in over-dense and under-dense environments are
significant at more than 90\% confidence level in all the redshift
bins up to $z=1.2$. We find that the best-fitting values of the LF
slope are systematically steeper in under-dense environments. We
measure a difference in the slope of the LF in under-dense versus
over-dense environments of $\Delta \alpha=0.28\pm 0.09$, $0.29 \pm
0.12$, $0.20 \pm 0.22$ in the redshift bins $z=$[$0.25-0.6$],
[$0.6-0.9$], [$0.9-1.2$], respectively. These differences are
significant at 3$\sigma$, 2$\sigma$ and 1$\sigma$, respectively. In
the redshift bin $z=$[$1.2-1.5$], the $\alpha-M^*$ parameters are too
weakly constrained to conclude on a variation of $\alpha$. In the
redshift bin $z=1.2-1.5$ the environmental dependency seems to be
diminishing, but this might be a result of missing faint and red
galaxies due to the VVDS selection function (see \cite
{Marinoni05}2005, \cite{Cucciati06}2006). Since error contours are
mostly degenerated along the magnitude axis, we observe no significant
difference in $M^*$ between under-dense and over-dense environment
over the redshits range $z=0.25-1.5$.  According to the 90\%
confidence contours, we conclude that the difference in $M^*$ between
under- and over-dense environments can not be greater than $\sim$1
mag, $\sim$0.8 mag and $\sim$1 mag at $z=0.25-0.6$, $z=0.6-0.9$ and
$z=0.9-1.2$, respectively.

\begin{figure*}
\centering \includegraphics[width=16cm]{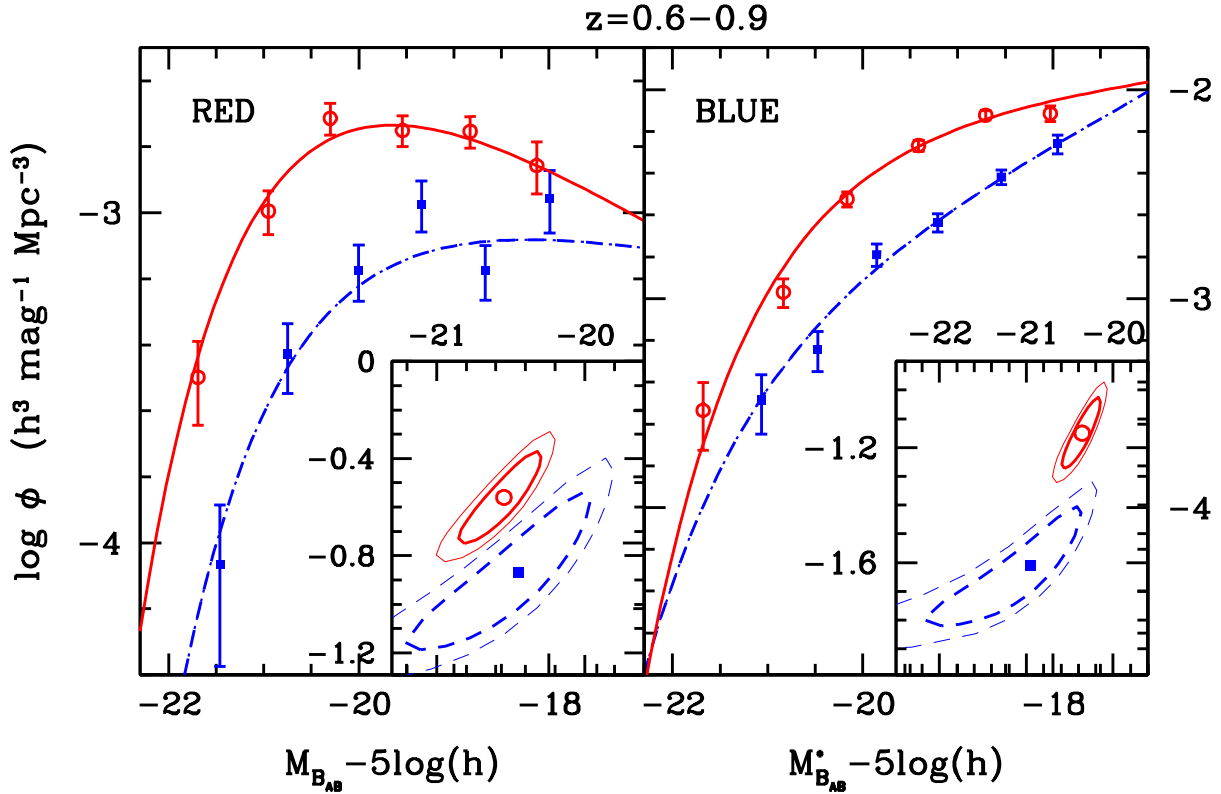}
\caption{B-band LF for under-dense (blue solid squares and dashed
  lines) and over-dense environment (red open circles and solid lines)
  at $z=0.6-0.9$ for red (left panel) and blue galaxies (right
  panel). Symbols are the same as in Fig.\,\ref{LFglob}.}
\label{LFtype}
\end{figure*}

Making the assumption that the LF shape ($M^*$ and $\alpha$) for a
given type is universal and does not depend on the environment, a
change in the relative fraction of each type as a function of the
environment could change the global LF shape. The large number of
galaxies at $z=0.6-0.9$ in the VVDS sample allows us to measure the LF
per type and environment simultaneously to test whether this
assumption is correct.  We split the sample at $z=0.6-0.9$ in red
$M_U-M_V>1.5$ and blue $M_U-M_V<1.5$ galaxies, which corresponds
roughly to the valley in the color bimodality of the VVDS data
(\cite{Franzetti06}2006).  Figure\,\ref{LFtype} shows the shape of the
LF for red and blue galaxies in under- and over-dense
environments. Independantly of the spectral type, the LF slope is
steeper in under-dense environments. Therefore, the excess of red
galaxies in over-dense environment shown in \cite{Cucciati06}(2006)
partially explains the environmental dependency of the global LF shape
since we find that even the type-specific LF significantly depends on
environment.

\begin{figure}
\centering \includegraphics[width=9.cm]{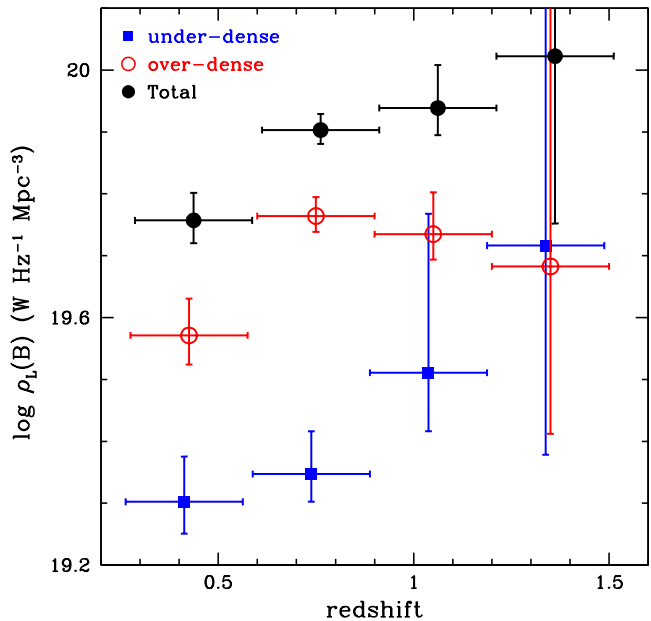}
\caption{Evolution of the Luminosity Density $\rho_L$ measured in the
  $B$ band as a function of redshift. Open red points refer to the
  over-dense environment, solid blue squares to the under-dense
  environment and solid black circles to the total sample.}
\label{LD}
\end{figure}

We measure the evolution of the luminosity density $\rho_L$ in the
rest-frame $B$ band. $\rho_L$ is directly derived from
$\rho_L=\int^{+\infty}_0 L \, \Phi(L) \, dL$ and we use the Schechter
parameters listed in Table \ref{tab1} to express $\Phi(L)$. Error bars
are based on the extreme values of $\rho_L$ calculated at each point
of the $\alpha-M^*$ 1$\sigma$ error contour. Fig.\,\ref{LD} shows the
evolution of the luminosity density $\rho_L$ as a function of redshift
in both environments. In under-dense environment, we find a continuous
increase of $\rho_L$ from $z=0.25-0.6$ up to $z=1.2-1.5$ by a factor
2.6. The LD evolves differently in over-dense environment: $\rho_L$
increases between $z=0.25-0.9$ then decreases or flattens at larger
redshifts to become lower or equivalent to the LD in under-dense
environment. An increase of the LD at the same rate than between
$z=0.25$ and $z=0.9$ is excluded at more than 3$\sigma$.  Tresse et
al. (2006, in preparation) show that the uncertainty on the slope is
the main source of uncertainty on $\rho_L$ at high redshift. Since the
error contours at $z=1.2-1.5$ reach $\alpha<-2$, the contribution of
the faint galaxies to the LD is not constrained. However, the
measurements of $\rho_L$ in the bin $z=1.2-1.5$ give a good indication
of the LD expected for a ``reasonable'' slope $\alpha \sim -1.5$ (e.g.
$\alpha=-1.44$ at $z=0.7-1$ for \cite{Poli03}2003, $\alpha=-1.6$ at
$z\sim 3$ for \cite{Steidel99}1999). At $z=1.2-1.5$, the decrease of
the LD is not constrained statistically since $\alpha$ is not
constrained. However, the slope $\alpha=-1.48$ used at $z=1.2-1.5$ is
steeper by $\Delta \alpha=0.26$ and $\Delta \alpha=0.42$ than the
slope measured at $z=0.6-0.9$ and $z=0.9-1.2$.  Therefore, using a
simple continuity argument for the slope between $z=0.9-1.2$ and
$z=1.2-1.5$, we do not expect an increase of the LD at $z>1.2$.

If we relate the observed evolution of the LD to the evolution of the
Schechter parameters, we find that the continuous increase of $\rho_L$
in under-dense regions is driven by a continuous brightening of $M^*$,
with $\Delta M^*_B\sim 0.6$ mag between $z=0.25$ and $z=1.5$. We
observe a similar brightening in over-dense environment with $\Delta
M^*_B\sim 0.8$ mag. The strong decrease of $\phi^*$ in over-dense
environment compensate the brightening of $M*$ and explains the
non-monotonic behavior of the LD observed at $z \sim 0.9$.

\section{Discussion}

We have found that the shape of the luminosity function shows a
significant dependency on the galaxy density contrast, and hence on
the environment, up to a redshift at least $z=1.2$. We find that the
values of the LF slope are systematically steeper by $\Delta \alpha
\sim 0.2-0.3$ in under-dense environments. We therefore detect at
these high redshifts an environmental dependency on the slope whereas
this dependency is observed only on $M^*$ in the local Universe (eg.
\cite{Marinoni99} 1999, \cite{Ramella99} 1999, \cite{DePropris02}2002,
\cite{Croton05}2005). Local surveys at $z\sim 0.1$ find a global LF
slope largely insensitive to the large-scale environment, which is
explained in terms of a steepening of the LF slope for early type
galaxies in over-dense regions (\cite{Croton05}2005). At high
redshifts $z=0.6-0.9$, we show that the LF of the red galaxy
population, instead, shows the opposite trend with a steepening of the
slope toward under-dense regions (Fig.\,\ref{LFtype}). Our results are
therefore suggestive of an increase of the population of faint red
objects in high density environments as a function of cosmic time,
consistent with the evolution of the color-environment relationship
observed by \cite{Cucciati06}(2006).  At variance with our findings on
the slope $\alpha$, we find that at redshifts up to $z\sim1.5$, $M^*$
is not significantly different in under- and over-dense environments,
but our data cannot exclude variations of a few tens of a magnitude
compatible with what has been found at low redshifts (e.g.  an
increase of $M^*$ of $\Delta M^*=0.3$ brighter in cluster environments
than in the field \cite{DePropris02}, 2002).

At $z\sim 1$ the luminosity distribution of galaxies that inhabit
low-density regions is observed to contain a large number of faint
galaxies, while the brighter galaxies are preferentially populating
high-density regions. The global trend for an excess of bright
galaxies in over-dense regions as shown in the local Universe
(\cite{Kauffmann04} 2004, \cite{Gomez03} 2003, \cite{Croton05}2005) is
therefore already present at $z\simeq 1$. This result is consistent
with the VVDS luminosity-dependant clustering results (\cite{Pollo06}
2006) and results from \cite{Bundy06}(2006). The mass function of dark
matter halos is expected to show a dependence on the large-scale
environment, as predicted using an extension of the Press-Schechter
formalism and confirmed in N-body cold dark matter simulations (e.g
\cite{Mo96}1996, \cite{Lemson99}1999).  It is predicted that dark
matter halos in over-dense regions should be more biased toward high
masses than those forming in lower density regions. Our result
therefore does not rule out that massive galaxies are build up from
merging but rather implies that merging should have been already very
efficient at $z >1$.

Using the LFs, we can directly compare the relative light emissivity
in under- and over-dense regions. We observe a continuous decrease of
the Luminosity Density (LD) in under-dense environments from $z=1.5$
up to $z=0.25$, by a factor 2.6, while the LD evolution in over-dense
environments presents a peak at $z\sim 0.9$. The continuous decrease
of the LD in under-dense environments from $z=1.5$ up to $z=0.25$ is
induced by the continuous fading of $M^*$ with cosmic time. Galaxies
in under-dense environments are expected to be located in more
isolated small size halos (\cite{Mo04}2004) and should therefore
evolve mainly with a passive consumption of the internal gas present
in their halos. The continuous decrease of the LD in under-dense
environments observed in our data could then be due to a continuous
decrease of the star formation rate in these more isolated systems.
The continuous fading of $M^*$ in over-dense environment from $z=1.5$
to $z=0.25$ should also lead to a continuous decrease of the LD in
over-dense environment. However, an increase of $\phi^*$ in over-dense
environments with cosmic time in effect compensates for this fading,
and instead is producing an increase of the LD between $z=1.5$ and
$z=0.9$. A relative change in the number of galaxies present above and
below $\delta=0$ could therefore explain the observed peak in the LD
for the over-dense regions. In the context of the standard
hierarchical scenario, the constant infalling of small-size halos
(which preferentially host faint galaxies) into larger dark matter
structures could explain an increase of galaxies in over-dense region.
Stated differently, since galaxies become more clustered with cosmic
time (e.g. \cite{LeFevre05b}2005b), the relative fraction of galaxies
in over-dense region increases with cosmic time. A complex interplay
between the star formation rate decreasing on the one hand and the
hierarchical growth of large-scale structures on the other hand could
explain the peak in the LD evolution at $z\sim 0.9$. Mechanisms like
strangulation resulting from the removal of the gas reservoir in
larger size halos (e.g.  \cite{Balogh00} 2000) or gas stripping by the
hot interstellar medium (\cite{Gunn72}1972) could also play a
significant role in quenching the star formation in infalling
galaxies, explaining the increasing number of faint red galaxies with
cosmic time.

\section{Summary}

We have measured the evolution of the galaxy Luminosity Function as a
function of large-scale environment up to $z=1.5$ from the VIMOS-VLT
Deep Survey (VVDS) first epoch data. The 3D galaxy density field is
reconstructed using a gaussian filter of smoothing length $5 h^{-1}$
Mpc and a sample of 6582 galaxies with $17.5 \le I_{AB} \le 24$ and
measured spectroscopic redshifts. We split the sample in four redshift
bins up to $z=1.5$ and classify galaxies depending on whether they are
located in under-dense or over-dense environments relative to the
average density contrast $\delta=0$. We compute the LF and LD for each
of these subsamples using our dedicated tool ALF.

We find that the LF shape, characterized with the Schechter parameters
$\alpha-M^*$, depends strongly on the large-scale environment from
$z=0.25$ up to $z=1.2$: the differences between the $\alpha-M^*$
best-fitting values in over-dense and under-dense environments are
significant at more than 90\% confidence level in all the redshift
bins up to $z=1.2$. We therefore conclude that either the shape of the
LF is imprinted very early on in the life of the Universe, a `nature'
process, or that `nurture' physical processes shaping up the
environment relation have already been efficient earlier than a
look-back time corresponding to 30\% of the current age of the
Universe.

The LF shape is observed to have a steeper slope in under-dense
environments. We find $\alpha=-1.32\pm 0.07, -1.35\pm 0.10, -1.42\pm
0.18$ in under-dense environment and $\alpha=-1.08\pm 0.05, -1.06\pm
0.06$, $-1.22\pm 0.12$ in over-dense environment in the redshift bins
$z=$[$0.25-0.6$], [$0.6-0.9$], [$0.9-1.2$], respectively using a
best-fit Schechter luminosity function. At variance, local
measurements at $z\sim 0.1$ (e.g. \cite{Croton05}2005) do not find any
dependency of the LF slope on the environment. We tentatively
interpret this change as an increase of the density of faint red
galaxies in over-dense environment along cosmic time.

Finally, we measure the evolution of the luminosity density $\rho_L$
in the rest-frame $B$ band. In under-dense environments, we find a
continuous increase of $\rho_L$ from $z=0.25-0.6$ up to $z=1.2-1.5$ by
a factor 2.6, driven by a continuous brightening of $\Delta M^* \sim
0.6$ mag from $z=0.25$ to $z=1.5$. This could be the result of a
passive evolution of the galaxies present in more isolated small size
halos. The LD evolves differently in over-dense environment: $\rho_L$
increases between $z=0.25-0.9$ then decreases or flattens at larger
redshifts. We interpret the peak at $z=0.9$ by a complex interplay
between the decrease of the star formation rate and the increasing
fraction of galaxies in over-dense environment due to the hierarchical
growth of structures.

\begin{acknowledgements}
  This research has been developed within the framework of the VVDS
  consortium. We thank the ESO staff at Paranal for their help in the
  acquisition of the data. This work has been partially supported by
  the CNRS-INSU and its Programme National de Cosmologie (France) and
  Programme National Galaxies (France), and by Italian Ministry (MIUR)
  grants COFIN2000 (MM02037133) and COFIN2003 (num.2003020150).  The
  VLT-VIMOS observations have been carried out on guaranteed time
  (GTO) allocated by the European Southern Observatory (ESO) to the
  VIRMOS consortium, under a contractual agreement between the Centre
  National de la Recherche Scientifique of France, heading a
  consortium of French and Italian institutes, and ESO, to design,
  manufacture and test the VIMOS instrument.

\end{acknowledgements}

\end{document}